\documentclass{article}

\usepackage{arxiv}

\usepackage{tikz}
\usepackage{enumitem}
\usepackage[utf8]{inputenc} 
\usepackage[T1]{fontenc}    
\usepackage{hyperref}       
\usepackage{url}            
\usepackage{booktabs}       
\usepackage{amsfonts}       
\usepackage{nicefrac}       
\usepackage{microtype}      
\usepackage{graphicx}
\usepackage[numbers]{natbib}
\usepackage{doi}
\usepackage{amsmath}
\usepackage{amssymb}
\usepackage{bm}
\usepackage{mathtools}

\title{System-Aware Contextual Digital Twin for ICS Anomaly Diagnosis}

\author{ \href{https://orcid.org/0009-0000-3621-5175}{\includegraphics[scale=0.06]{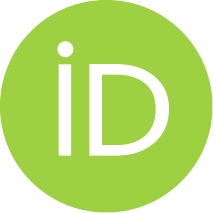}\hspace{1mm}Eungyu Woo} \\
	Department of EECS\\
	DGIST\\
	Daegu, South Korea \\
	\texttt{ekwoo@dgist.ac.kr} \\
	\And
	\href{https://orcid.org/0009-0002-1687-7432}{\includegraphics[scale=0.06]{orcid.pdf}\hspace{1mm}Yooshin Kim} \\
	Department of EECS\\
	DGIST\\
	Daegu, South Korea \\
	\texttt{yooshin0303@dgist.ac.kr} \\
	\And
	\href{https://orcid.org/0000-0003-1753-8977}{\includegraphics[scale=0.06]{orcid.pdf}\hspace{1mm}Wonje Heo} \\
	Department of EECS\\
	DGIST\\
	Daegu, South Korea \\
	\texttt{heo\_wonje@dgist.ac.kr} \\
	\And
	\href{https://orcid.org/0009-0004-7088-3846}{\includegraphics[scale=0.06]{orcid.pdf}\hspace{1mm}Donghoon Shin} \\
	Department of EECS\\
	DGIST\\
	Daegu, South Korea \\
	\texttt{dshin@dgist.ac.kr}
}

\renewcommand{\shorttitle}{SCDT}

\hypersetup{
pdftitle={A template for the arxiv style},
pdfsubject={q-bio.NC, q-bio.QM},
pdfauthor={Eungyu Woo, Yooshin Kim, Wonje Heo, Donghoon Shin},
pdfkeywords={Large Language Model, Industrial Control System, Contextual Digital Twin, Anomaly Detection},
}

\begin{document}

\maketitle

\begin{abstract}
	Industrial Control Systems (ICS) integrate computing, physical processes, and communication to operate critical infrastructures such as power grids, water treatment plants, and oil and gas facilities. As ICS become increasingly targeted by cyberattacks, timely and reliable anomaly diagnosis is essential for protecting operational safety. However, existing ICS anomaly detection approaches face practical limitations: supervised methods require extensive labeled attack data and suffer from class imbalance, while model-based detectors often lack the ability to provide deep insight into the root causes of anomalies, leading to elevated false alarms and making it difficult for operators to initiate a timely response.
In this work, we propose a system-aware unsupervised framework for ICS anomaly diagnosis that combines lightweight online detection with contextual explanation. The system identifies deviations from observed normal behaviors without prior knowledge of system topology. To support actionable response, we further concatenate a contextual digital twin augmented with an Large Language Model (LLM) to enhance interpretability, which translates detection evidence into grounded diagnostic hypotheses and verification steps for operators.
Experiments on public ICS benchmarks demonstrate that the proposed framework achieves real-time detection efficiency and provides consistent, interpretable anomaly diagnoses, enabling low-latency warning and practical deployment in complex industrial environments.
\end{abstract}

\keywords{Large Language Model \and Industrial Control System \and Contextual Digital Twin \and Anomaly Detection}

\section{Introduction}
Industrial Control Systems (ICS) operate critical infrastructures such as water treatment plants, power grids, and manufacturing facilities~\cite{nist80082r3,CPSsurvey}. Their continuous and reliable operation is essential for public safety and economic stability~\cite{CPSsurvey}. In recent years, however, ICS have frequently been targeted by cyberattacks, ranging from stealthy process manipulation to disruptive sabotage~\cite{Blackenergy,stuxnet}. Unlike conventional information technology (IT) systems, failures in ICS can directly translate into physical consequences, making timely and reliable anomaly detection a core requirement for safe operation~\cite{nist80082r3}.

Despite extensive research on ICS anomaly detection, existing methods still struggle to bridge the gap between reliable detection and practical diagnosis~\cite{10.1145/3609333,mr2021machine}. Supervised methods rely on labeled attack data~\cite{koay2023machine}, but in practice such data are extremely scarce, highly imbalanced, and quickly outdated as attack strategies evolve, which limits their robustness in real deployments~\cite{SHYAA2024109143}. Unsupervised methods avoid label dependence and are therefore more suitable for realistic settings~\cite{koay2023machine}, but they typically focus on detecting statistical deviations without providing sufficient insight into the underlying causes~\cite{10.1145/3460120.3484589, 10.1145/3600160.3600193}. As a result, even when anomalies are detected, operators still face a time-consuming diagnostic process, inspecting raw signals and logs to identify the underlying causes and system-level implications~\cite{XAI1}. This gap motivates the need for unsupervised anomaly detection frameworks that are not only sensitive to abnormal behavior but also capable of supporting explanation and diagnosis~\cite{mr2021machine}.

\begin{figure}[t]
  \centering
  \includegraphics[width=0.5\linewidth]{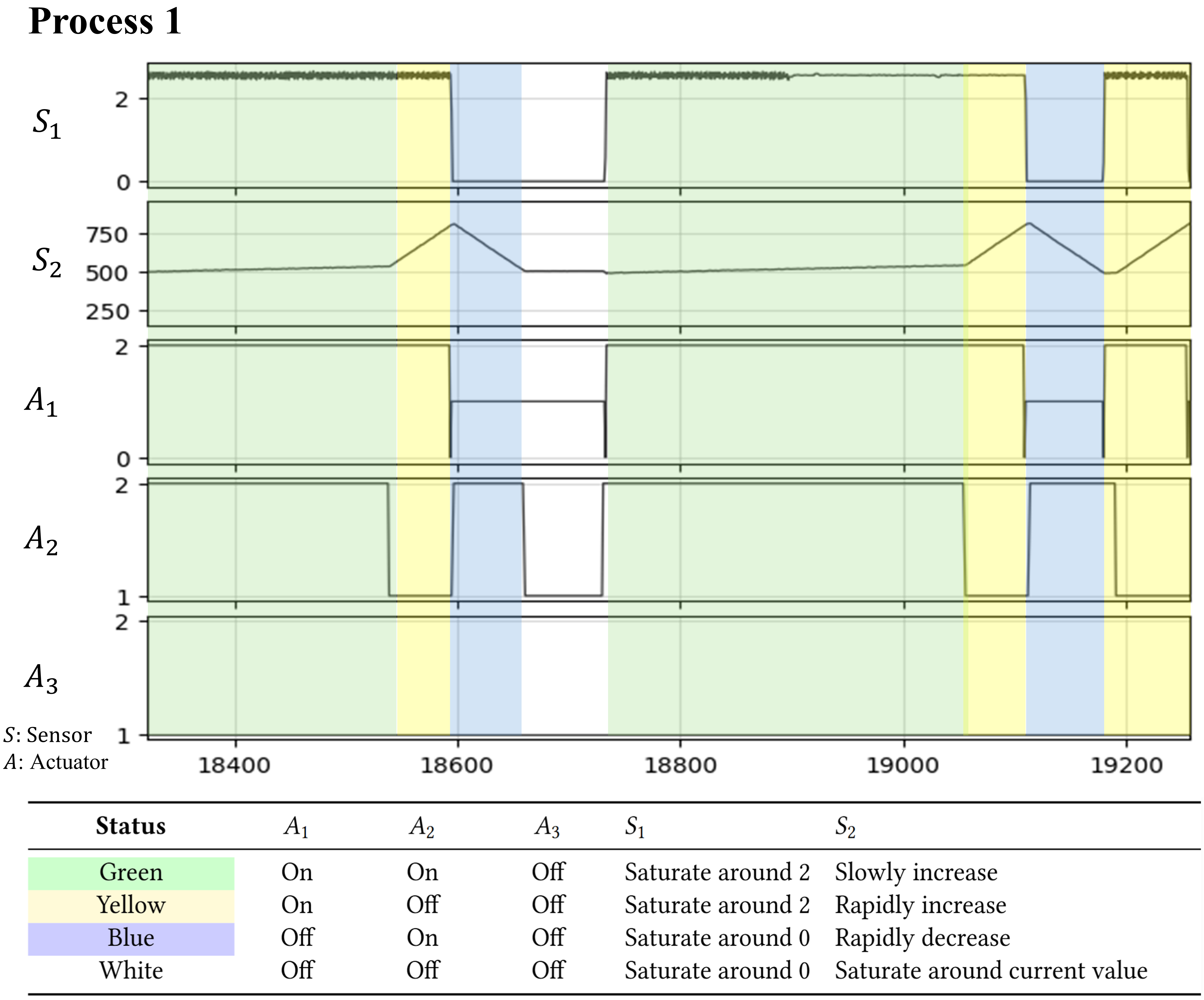}
  \caption{An illustration of Process 1 in the SWaT testbed~\cite{SWAT}. Colored regions indicate recurrent operational contexts, each characterized by a specific actuator configuration and a consistent pattern of sensor evolution. The table summarizes the context-dependent actuator states and corresponding sensor behaviors, revealing invariant relationships between actuators and sensors under the same operational objective.}
  \label{fig:swat_process_1}
\end{figure}

A defining characteristic of ICS is the structured interaction between actuators and sensors~\cite{10.1145/3453155, DT9}. Operators (or control logic) manipulate actuators to achieve specific operational objectives, sensors observe the resulting system states, and the actuators are subsequently adjusted based on this feedback, forming a closed control loop. Through this cyclic process, stable relationships emerge between actuator configurations and sensor responses. In real-world testbeds such as water treatment systems~\cite{SWAT}, we observe that under a fixed operational objective (e.g., maintaining tank levels in Process~1), similar actuator configurations repeatedly induce similar patterns of sensor evolution (Fig.~\ref{fig:swat_process_1}). This regularity indicates the existence of invariant rules, or more broadly, operational contexts, governing sensor-actuator relationships. These context-dependent invariants provide a natural basis for modeling expected system behavior under normal operation, beyond mere pointwise signal statistics~\cite{DT6, ZHU2025106164}.

In practice, human operators diagnose anomalies by reasoning over system context~\cite{fung2025adopting}: they interpret the current system behavior with respect to the expected behavior under the assumed operating regime and identify where this contextual consistency is violated. This context-driven reasoning process is critical for rapid recovery, yet it relies heavily on scarce domain expertise~\cite{SANS2025StateOfICSSecurity}. When diagnosis is delayed, the consequences can escalate into human, social, and physical harm, highlighting the need for effective decision-support mechanisms~\cite{nist80082r3}. In this regard, large language models (LLMs) have recently attracted attention as potential aids for complex reasoning tasks~\cite{LLMAD, zhang-etal-2025-ufo, 11112952, 10.5555/3666122.3666983}. However, directly applying LLMs to raw numerical time-series data in ICS is inappropriate due to the risk of hallucination and the absence of proper grounding~\cite{zeng2026halluguard, 10.1145/3703155, 10.24963/ijcai.2024/921}. Instead, structured, context-aware, and evidence-based diagnostic hypotheses are required to bridge numerical detection with semantic reasoning~\cite{jin2024timellm}.

In this work, we propose SCDT (System-aware Contextual Digital Twin), a framework that reconceptualizes ICS anomaly diagnosis as a two-stage, evidence-grounded reasoning problem.  In the first stage, SCDT performs fast unsupervised system-status monitoring by learning context-conditioned behavioral envelopes from historical normal operation. This captures invariant regularities of ICS behavior across diverse operating regimes without relying on attack labels. During online monitoring, SCDT evaluates whether the observed system state is compatible with these learned envelopes and triggers a system-level alert upon decisive deviations. When the operational context cannot be matched reliably, SCDT explicitly marks the case as ambiguous rather than forcing an arbitrary decision, preserving safety and enabling downstream reasoning.

In the second stage, SCDT concatenates a contextual digital twin powered by an LLM to support diagnosis. The LLM operates on structured detector evidence—such as violated behavioral dimensions, deviation margins, and the matched operational context—together with system metadata to generate grounded diagnostic hypotheses and verification steps. This separation preserves evidence-based safety-critical detection, while leveraging LLMs for semantic interpretation and operator-facing explanations.


The key contributions are summarized as follows:
\begin{itemize}[leftmargin=10pt, itemindent=0pt, itemsep=0pt, topsep=0pt]
    \item \textbf{System-aware unsupervised diagnosis.} We formulate ICS anomaly diagnosis around sensor-actuator context, enabling context-conditioned monitoring aligned with closed-loop control.
    \item \textbf{LLM-grounded contextual explanations.} We introduce an LLM-based diagnostic layer that converts detector evidence into actionable, grounded hypotheses without altering the detector's decisions.
    \item \textbf{Robust handling of ambiguous contexts.} We learn reusable normal envelopes across diverse operating regimes and explicitly treat low-confidence context matching as ambiguity.
    \item \textbf{Real-time, deployment-oriented design.} We provide lightweight online screening with system-level aggregation to reduce false alarms under legitimate regime changes.
\end{itemize}

\section{Related Work}
\paragraph*{\textbf{Multivariate Time Series Anomaly Detection}}
Unsupervised anomaly detection for multivariate time series (MTS) is typically based on distance/subspace methods or deep models trained with reconstruction or prediction objectives. Classical baselines such as \emph{PCA} \cite{PCA} and \emph{KNN} \cite{KNN} flag anomalies using projection error or neighbor distance, and \emph{OCSVM} \cite{OCSVM} learns a boundary that encloses normal data. These methods are simple and fast, but they mainly rely on global statistics and are sensitive to changes in operating modes.
Reconstruction-based deep models use reconstruction error as an anomaly score. \emph{AE} \cite{AE} is the standard baseline, and \emph{DAGMM} \cite{DAGMM} combines an autoencoder with mixture modeling to compute energy scores. For sequential dynamics, \emph{LSTM-VAE} \cite{LSTMVAE} uses recurrent encoder-decoder structures. Adversarial variants such as \emph{MADGAN} \cite{MADGAN} and \emph{USAD} \cite{USAD} increase the separation between normal and abnormal samples by training generator/discriminator-style components. These models can capture complex patterns, but the score is still driven by how well the model fits the data, which often does not reflect control context.
Prediction-based and structure-aware approaches incorporate cross-sensor dependencies. \emph{GDN} \cite{GDN} learns sensor relations and detects anomalies from forecasting residuals, and \emph{FuSAGNet} \cite{FuSAGNet} models temporal dynamics and feature subspaces with attention and recurrent units. While these methods use correlations across variables, they still treat normality as statistical regularity of signals. In ICS, normality depends on actuator settings, process stage, and control intent, so the same sensor behavior can be normal in one context and abnormal in another \cite{DT6, DT9}. Many MTS methods do not explicitly represent this dependency, and consequently provide limited interpretability: they typically return anomaly scores without an explicit rationale linking the decision to actionable control-state explanations.

\paragraph*{\textbf{Digital Twins and LLM-Based Reasoning for ICS}}
High-fidelity digital twins (DTs) can support prediction and diagnosis by simulating plant dynamics, but they are costly to build and maintain \cite{DT1}. They require domain modeling, and understanding their internal logic often needs specialized tools \cite{DT2, DT3}. Updating DTs for changes in plant configuration is also expensive \cite{DT4}. Explainable AI has been used to improve DT transparency, but research in this area remains limited \cite{DT5}. These approaches identify which sensors are anomalous but fail to explain why the behavior violates the current control logic or actuator context. This creates demand for DT approaches that are easier to build, easier to update, and interpretable by design.
Interpretable approaches instead mine invariants or rules connecting actuator states and sensor responses \cite{DT6, DT9}. These methods can provide explicit constraints, but they often depend on handcrafted features and static rule sets, which are expensive to curate and do not adapt well to changes. To overcome these rigidities, Large Language Models (LLMs) have emerged as a promising alternative, offering flexible capabilities for contextual reasoning on structured logs and telemetry. Recent works utilize LLMs to automate invariant extraction from documentation \cite{DT7} and generating attack patterns for testing \cite{DT8}. However, current LLM-based approaches do not provide a method to continuously derive and maintain an explicit, evolving set of sensor-actuator rules that can be used for anomaly reasoning and human-readable explanations.

\begin{figure*}[ht]
  \centering
  \includegraphics[width=0.95\linewidth]{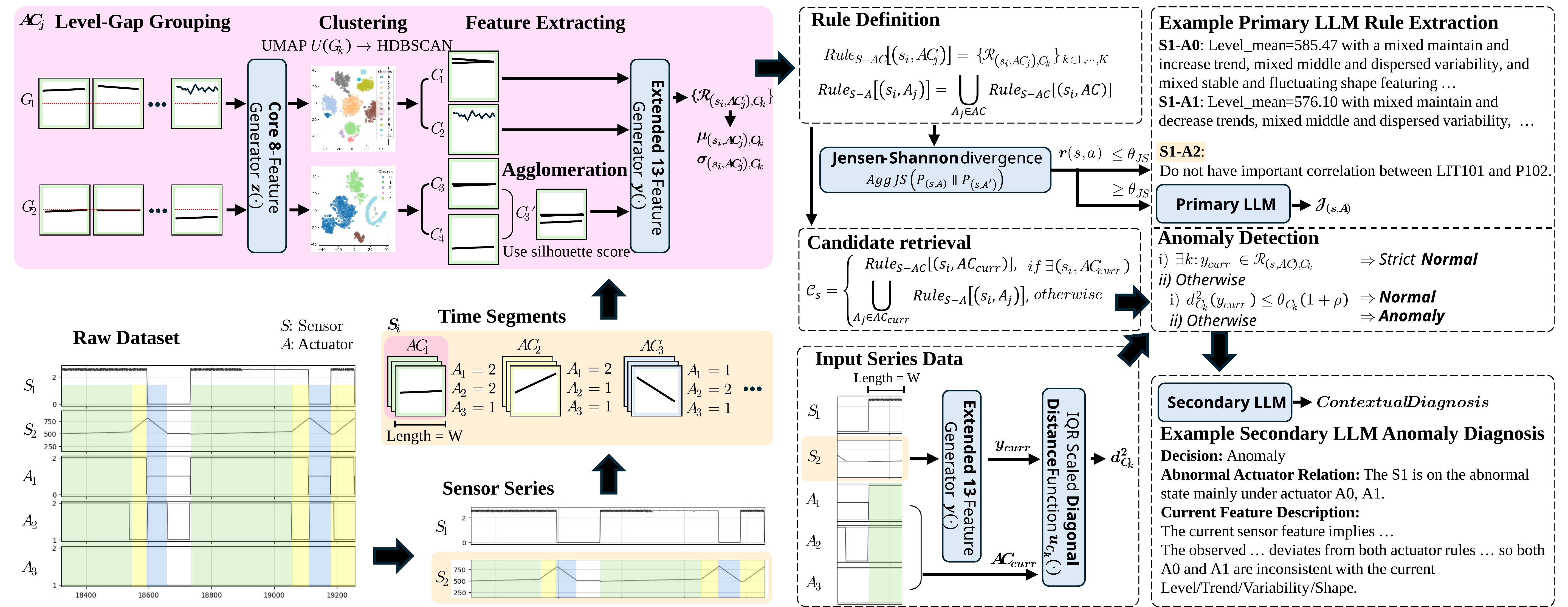}
  \caption{Overview of the proposed SCDT framework. The left panel shows training, where normal features are extracted and multi-stage S-AC clustering forms rule prototypes. The right panel shows inference, where the Primary LLM summarizes rules and the Secondary LLM detects anomalies from the current window by matching features to the rule set.}
  \label{fig:framework_overview}
\end{figure*}

\begin{figure}[ht]
  \centering
  \includegraphics[width=0.5\linewidth]{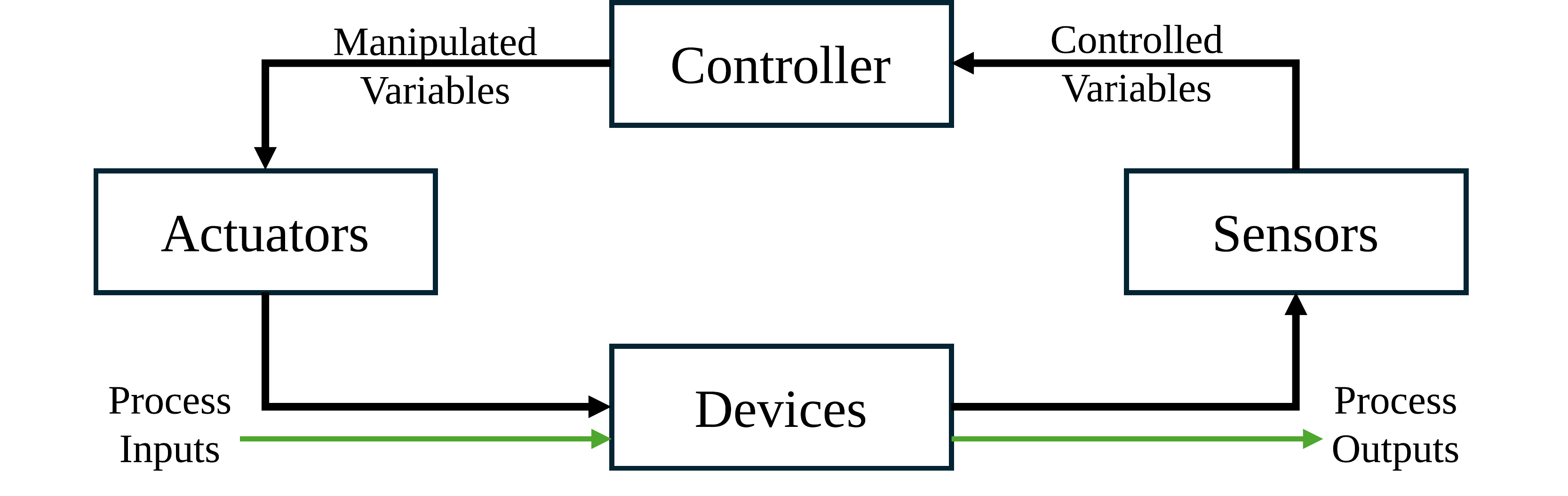}
  \caption{ICS Operation}
  \label{ICSOp}
\end{figure}

\begin{table*}[ht]
\centering
\caption{Feature taxonomy for window-level representation (13 features). Each feature contributes to characterize level, shape, trend, variability, or periodicity of ICS sensor behavior.}
\label{tab:taxonomy}
\begin{tabular}{lccccl}
\toprule
\textbf{Feature} & \textbf{Level} & \textbf{Variability} & \textbf{Shape} & \textbf{Trend} & \textbf{Description} \\
\midrule
$f_{\min}$                 & O & - & - & - & Window minimum (lower level bound) \\
$f_{\max}$                 & O & - & - & - & Window maximum (upper level bound) \\
$f_{\mathrm{mean}}$        & O & - & - & - & Average operating level \\
$f_{\mathrm{amp}}$         & O & O & - & - & Robust amplitude (0.995–0.005 quantile range) \\
$f_{\mathrm{std}}$         & - & O & O & - & Global variability (dispersion) \\
$f_{\mathrm{rmse}}$        & - & O & O & - & Deviation from linear trend (linearity residual) \\
$f_{\Delta}$               & - & - & O & - & Magnitude of local changes (first difference) \\
$f_{\Delta^2}$             & - & - & O & - & Sharpness or curvature of changes (second difference) \\
$f_{\mathrm{neg}}$         & - & - & O & - & Count of large negative jumps (drain/reset behavior) \\
$f_{\mathrm{spec}}$        & - & - & O & O & Non-DC spectral energy ratio (periodicity strength) \\
$f_{s,1}$                  & - & - & - & O & Early-window trend (local slope) \\
$f_{s,2}$                  & - & - & - & O & Mid-window trend (local slope) \\
$f_{s,3}$                  & - & - & - & O & Late-window trend (local slope) \\
\bottomrule
\end{tabular}
\end{table*}

\section{System-Aware Contextual Digital Twin (SCDT)}
\subsection{Overview}

In ICS, sensor trajectories are inherently \emph{context-dependent}; a specific sensor pattern may be normal under one actuator configuration yet abnormal under another. Figure~\ref{ICSOp} illustrates the closed-loop architecture, in which sensors, controllers, actuators, and physical processes interact through continuous feedback. To address this, we define the \emph{Sensor-Actuator Combination (S-AC)} context as the joint actuator state observed at the beginning of a sensor window.

The full SCDT pipeline, illustrated in Figure~\ref{fig:framework_overview}, is designed to model these contexts. Starting with window-level feature extraction, the system identifies normal operational modes via multi-stage clustering and encodes them into interpretable rule envelopes for fast screening. A Primary LLM translates these numerical prototypes into semantic expectations. At inference time, the system performs a hierarchical diagnosis: it first applies a deterministic check using the retrieved rules, and then employs a Secondary LLM to either explain definitive results or arbitrate ambiguous cases. This design effectively balances high-speed detection with deep, human-readable diagnostics.

\subsection{Statistical Feature Extraction for grounded normal state}

\paragraph*{Notation}
Let $x=\{x_0,\ldots,x_{W-1}\}$ denote a sensor window of length $W$, and let
$AC$ be the actuator combination at the window start. We denote the core
feature vector by $\mathbf{z}\in\mathbb{R}^{8}$ and the rule descriptor by
$\mathbf{y}\in\mathbb{R}^{13}$.

For actuator-level indexing, each actuator value is discretized into a finite state
$A_a \in \{0,\ldots,B\}$ by a fixed quantization (binning) map $A_a=b_a(\cdot)$.
With a slight abuse of notation, we refer to this discretized actuator state simply as $A$.

\paragraph*{Window-level feature extraction}
For each window $x=\{x_t\}_{t=0}^{W-1}$, define $\Delta x_t=x_{t+1}-x_t$ and
$\Delta^2 x_t=\Delta x_{t+1}-\Delta x_t$. We compute eight core statistics:
\begin{itemize}
    \item $f_{\mathrm{mean}}$: sample mean of $x$,
    \item $f_{\mathrm{amp}}$: robust amplitude $Q_{0.995}(x)-Q_{0.005}(x)$,
    \item $f_{\mathrm{std}}$: sample standard deviation,
    \item $f_{\mathrm{rmse}}$: RMSE to least-squares line $a t+b$,
    \item $f_{\Delta}$: mean absolute first difference $\tfrac{1}{W-1}\sum_t|\Delta x_t|$,
    \item $f_{\Delta^2}$: mean absolute second difference $\tfrac{1}{W-2}\sum_t|\Delta^2 x_t|$,
    \item $f_{\mathrm{neg}}$: number of negative jumps satisfying $\Delta x_t < T$.
    \item $f_{\mathrm{spec}}$: non-DC spectral ratio $\sum_{k\ge1}|X_k|^2 / (\sum_{k\ge0}|X_k|^2+\epsilon)$,
\end{itemize}
where $T=-5\cdot\mathrm{MAD}(\Delta x)$, 
$(a t+b)$ is the linear fit of $x$, and $X_k$ is the rFFT of $x-\mathrm{mean}(x)$.  
The core feature vector is
\[
\mathbf{z}=[f_{\mathrm{mean}},f_{\mathrm{amp}},f_{\mathrm{std}},f_{\mathrm{rmse}},f_{\Delta},f_{\Delta^2},f_{\mathrm{neg}},f_{\mathrm{spec}}].
\]

For rule construction and diagnosis, we extend the descriptor to 13 dimensions:
\[
\mathbf{y}=
[f_{\min}, f_{\max}, f_{s,1}, f_{s,2}, f_{s,3}, \mathbf{z}],
\]
where $f_{\min}=\min(x)$, $f_{\max}=\max(x)$, and $f_{s,i}$ is the slope of the
$i$-th third segment of the window (early/mid/late trend).

\paragraph*{Interpretable meta-attributes and the 8D/13D split}
We distinguish the 8-D core feature vector $\mathbf{z}$ from the 13-D extended descriptor $\mathbf{y}$
for both efficiency and interpretability.
During rule learning and clustering, we primarily use $\mathbf{z}$ to reduce computational overhead,
while handling absolute operating-band differences through a separate \emph{level-gap} margin rule; hence,
explicit extrema and segment-wise slopes are not required in the clustering space.
In the Table~\ref{tab:taxonomy}, for diagnosis and explanation, we switch to $\mathbf{y}=[f_{\min}, f_{\max}, f_{s,1}, f_{s,2}, f_{s,3}, \mathbf{z}]$
to provide operator-facing summaries in four meta-attributes:
\textbf{Level} via $(f_{\min}, f_{\max}, f_{\mathrm{mean}})$ (optionally complemented by the robust range $f_{\mathrm{amp}}$),
\textbf{Trend} via $(f_{s,1}, f_{s,2}, f_{s,3})$,
\textbf{Variability} via $(f_{\mathrm{std}}, f_{\mathrm{rmse}}, f_{\mathrm{amp}})$, and
\textbf{Shape} via $(f_{\Delta}, f_{\Delta^2}, f_{\mathrm{neg}}, f_{\mathrm{spec}})$.
This separation allows the system to focus on core dynamics during clustering while leveraging the full feature set to capture the complete shape of normal states during inference, leading to superior detection performance.

\subsection{Multi-stage Rule Clustering}
\paragraph*{Clustering for each S-AC}
For each sensor $s$ and actuator combination $AC$, we collect the set of core
features $\{\mathbf{z}_j\}_{j=1}^{n}$ extracted from windows observed under
$(s,AC)$, where $n$ denotes the total number of time windows. Because normal operation under a fixed $(s,AC)$ can still include
multiple modes (e.g., different inflow regimes or upstream conditions), we
identify modes via the following multi-stage procedure:

(i) Level-gap grouping. We first partition windows into coarse level
groups $\{G_i\}$ using level-related statistics (e.g., $f_{\min}, f_{\max}$) so
that windows with clearly different operating levels are not clustered together.
Each group yields a matrix $\mathbf{Z}^{(i)}\in\mathbb{R}^{n_i\times 8}$.

(ii) Group-wise UMAP \cite{UMAP}. UMAP efficiently captures non-linear correlations and preserves global structure even at high speeds, which is crucial for distinguishing complex operational modes. For each group $G_i$, we embed
$\mathbf{Z}^{(i)}$ into 2D:
\[
\mathbf{Z}^{(i)}\in\mathbb{R}^{n_i\times 8}
\;\mapsto\;
\mathbf{U}^{(i)}\in\mathbb{R}^{n_i\times 2}.
\]
This stabilizes clustering under heterogeneous regimes by preserving local
neighborhoods within each level group.

(iii) Group-wise HDBSCAN \cite{HDBSCAN}. We apply HDBSCAN on each $\mathbf{U}^{(i)}$ to
discover density-separated clusters. By leveraging cluster stability measures across a hierarchy, HDBSCAN automatically determines density thresholds without requiring a manual global $\epsilon$, making it robust to varying cluster densities. Single-cluster outcomes are allowed.

(iv) Global consolidation. We aggregate clusters from all groups under
the same $(s,AC)$. Since HDBSCAN may yield an excessive number of fine-grained clusters, we merge similar clusters by optimizing the silhouette score \cite{Silhouette} to ensure that each resulting cluster represents a distinct operational mode.
The final clusters $\mathcal{C}_{(s,AC)}=\{C_1,\ldots,C_K\}$ represent normal
modes under the given context.

\paragraph*{Rule construction}
For each mode $C_k\in\mathcal{C}_{(s,AC)}$, we collect the corresponding 13D
descriptors $\{\mathbf{y}\}$ and define an element-wise quantile envelope:
\[
\mathcal{R}_{(s,AC),k}=\big[Q_{\alpha}(\mathbf{y}),\,Q_{1-\alpha}(\mathbf{y})\big],
\]
where $\alpha$ is small (e.g., $\alpha=0.005$). This envelope compactly encodes
admissible feature ranges for that mode.

\paragraph*{Fast robust distance model}
To support fast screening against many candidate modes, we additionally store a
robust diagonal distance \cite{Mahalanobis} model per cluster in a robustly scaled space.

Let $\tilde{\mathbf{y}}$ denote $\mathbf{y}$ after a robust per-key scaling
(e.g., robust scaler fitted on all samples of the key). For each cluster $C_k$,
we compute a robust center $\boldsymbol{\mu}_k$ (median) and a robust scale
vector $\boldsymbol{\sigma}_k$ (IQR with a safety floor to avoid degenerate
scales). For an input $\tilde{\mathbf{y}}$, we define
\[
\mathbf{u}_k = (\tilde{\mathbf{y}}-\boldsymbol{\mu}_k)\oslash(\boldsymbol{\sigma}_k+\epsilon),
\qquad
\mathbf{u}_k \leftarrow \mathrm{clip}(\mathbf{u}_k, -z_{\max}, z_{\max}),
\]
and define the diagonal robust distance, where $\oslash$ denotes element-wise division.
\[
d_k^2(\tilde{\mathbf{y}})=\sum_{j=1}^{13} u_{k,j}^2.
\]
We calibrate a cluster-specific threshold $\theta_k$ from training distances
(e.g., $\theta_k = Q_{q}(d_k^2)$ with a high $q$ such as $0.999$), and apply a
baseline floor for small clusters to prevent overly tight acceptance.

\subsection{LLM-Driven Rule Diagnosis}
\paragraph*{Primary-LLM semantic rule generation}
For each mode rule $\mathcal{R}_{(s,AC),k}$, a frozen Primary LLM generates a short natural-language description that summarizes expected behavior under the corresponding actuator context (e.g., expected level/trend/variability/shape patterns). To obtain this description, we prompt the LLM with the feature statistics of each cluster and request a concise, single-sentence characterization. In particular, \emph{Level} is represented by the cluster mean as a central tendency, with the admissible range determined by $(f_{\min},f_{\max})$, and further refined using the robust amplitude $f_{\mathrm{amp}}$ to capture typical within-window variation. \emph{Trend} is inferred from the three segment-wise slopes $(f_{s,1},f_{s,2},f_{s,3})$ and mapped to one of {\texttt{increase}, \texttt{decrease}, \texttt{maintain}} to describe the overall direction across the window. \emph{Variability} is categorized into {\texttt{compact}, \texttt{middle}, \texttt{dispersed}} based on dispersion-related features (e.g., $f_{\mathrm{std}}$ and robust difference statistics), and \emph{Shape} is summarized as {\texttt{stable}, \texttt{middle}, \texttt{fluctuating}} using higher-order dynamics and spectral cues (e.g., $\Delta$- and $\Delta^2$-based statistics, negative jumps, and the non-DC spectral ratio). After computing these attributes for each cluster associated with a rule, we select the dominant cluster modes (i.e., those with the highest support) and record their LLM-generated sentences as the semantic rule text stored alongside the numerical envelopes for downstream diagnosis.

\paragraph*{Actuator-level marginalization and sensor-actuator relatedness (S-A)}
To generalize beyond exact S-AC keys and to enable actuator-aware diagnosis, we build an actuator-level index over discretized actuator states.
Let $A_a=b_a(AC)\in\{0,\ldots,B\}$ denote the discretized state of actuator $a$ at the window start under context $AC$.
For each sensor $s$, actuator $a$, and actuator state $A$, we pool normal modes across all S-AC contexts whose actuator state matches $A$:
\[
\mathcal{C}_{(s,a,A)} \;=\!\!\!\bigcup_{AC:\,b_a(AC)=A}\!\!\!\mathcal{C}_{(s,AC)},
\]
\[
\text{Rule}_{S\text{-}A}[(s,a,A)] \;=\; \{\mathcal{R}_{(s,AC),k}\;:\; C_k\in\mathcal{C}_{(s,a,A)}\}.
\]
Not all actuators influence a given sensor.
We therefore estimate a relatedness score $r(s,a)$ by measuring how the sensor's mode-summary distribution changes across actuator states.
For each $(s,a,A)$ we form a discrete signature of each pooled mode (e.g., tags derived from $\mathbf{y}$ capturing level/trend/variability/shape),
construct the empirical distribution $P_{s,a,A}$ over signatures, and define
\[
r(s,a)=\mathrm{Agg}_{A\neq A'}\mathrm{JS}\big(P_{s,a,A}\,\|\,P_{s,a,A'}\big),
\]
where $\mathrm{JS}$ is Jensen-Shannon divergence \cite{JShannon} and $\mathrm{Agg}$ is an average or maximum across state pairs.
Actuators with low $r(s,a)$ are treated as unrelated and excluded from LLM prompts.

\paragraph*{Primary-LLM actuator-state semantic rule bank}
For each related pair $(s,a)$ and each actuator state $A$, a frozen Primary LLM generates a 1-2 sentence expectation $\mathcal{T}_{s,a,A}$ that summarizes typical behavior under $\text{Rule}_{S\text{-}A}[(s,a,A)]$.
The prompt provides representative pooled modes.
The resulting texts $\{\mathcal{T}_{s,a,A}\}$ are stored for fast retrieval during diagnosis and arbitration.

\paragraph*{Fallback retrieval via S-A index}
If an exact $(s,AC)$ key is unavailable at inference, we fall back to the S-A index over discretized actuator states.
Given $AC_{\mathrm{curr}}$, we retrieve actuator-conditioned candidates for the current states:
\[
\widetilde{\mathcal{C}}_s \;=\; \displaystyle\bigcup_{a\in AC_{\mathrm{curr}}}\text{Rule}_{S\text{-}A}\big[(s,a,b_a(AC_{\mathrm{curr}}))\big].
\]
This retrieval provides both numerical candidates (envelopes/distances) and semantic expectations $\mathcal{T}_{s,a,b_a(AC_{\mathrm{curr}})}$ even when the full actuator combination is unseen.

\subsection{Inference Process}

\paragraph*{Candidate retrieval}
Given a live window, we compute $\mathbf{y}_{\mathrm{curr}}$ and observe
$AC_{\mathrm{curr}}$. For each sensor $s$, candidate mode rules are retrieved as
\[
\mathcal{C}_s=
\begin{cases}
\text{Rule}_{S\text{-}AC}[(s,AC_{\mathrm{curr}})], &
\text{if } (s,AC_{\mathrm{curr}})  \in \text{Rule},\\[4pt]
\displaystyle\bigcup_{a\in AC_{\mathrm{curr}}}\text{Rule}_{S\text{-}A}\big[(s,a,b_a(AC_{\mathrm{curr}}))\big], &
\text{otherwise}.
\end{cases}
\]
If $\mathcal{C}_s$ is empty, the sample is marked as \emph{ambiguous} and deferred
to LLM arbitration.

\paragraph*{Deterministic screening with envelopes and robust distances}
For each candidate mode $k\in\mathcal{C}_s$, we perform two rapid validation:

(i) Envelope membership.
If $\mathbf{y}_{\mathrm{curr}}$ lies inside the quantile envelope of any mode,
it is accepted as normal:
\[
\exists k:\;\mathbf{y}_{\mathrm{curr}}\in \mathcal{R}_{(s,AC),k}
\;\Rightarrow\; \text{Normal}.
\]

(ii) Robust distance acceptance.
Otherwise, we compute $d_k^2(\tilde{\mathbf{y}}_{\mathrm{curr}})$ and compare it
to an expanded threshold:
\[
d_k^2(\tilde{\mathbf{y}}_{\mathrm{curr}})\le \theta_k(1+\rho)
\;\Rightarrow\; \text{Normal},
\]
where $\rho\ge 0$ is a small relaxation factor that controls tolerance.

We define the per-mode \emph{margin} as
\[
m_k = \theta_k(1+\rho) - d_k^2(\tilde{\mathbf{y}}_{\mathrm{curr}}),
\]
and select the representative mode by $k^\star=\arg\max_k m_k$.

\paragraph*{Secondary-LLM diagnosis and arbitration with S-A semantics}
SCDT uses a frozen Secondary LLM in two modes.

(i) \emph{Diagnosis mode} When the detector yields a definitive decision (normal/anomaly), the Secondary LLM provides a context-aware explanation without altering the decision.
It receives (i) the current actuator states $b_a(AC_{\mathrm{curr}})$, (ii) the detector output, (iii) the top related actuators for the sensor ranked by $r(s,a)$, and (iv) the corresponding Primary-LLM expectations $\mathcal{T}_{s,a,b_a(AC_{\mathrm{curr}})}$ together with a compact deviation summary of $\mathbf{y}_{\mathrm{curr}}$.
It outputs a short diagnostic narrative identifying which actuator-sensor expectation is most inconsistent with the observation.

(ii) \emph{Arbitration mode} If deterministic screening is ambiguous (e.g., unseen $(s,AC_{\mathrm{curr}})$ and no confident acceptance among pooled candidates),
the Secondary LLM also returns the final normal/anomaly decision.
In this case, it bases its judgement on the same S-A expectations, comparing $\mathbf{y}_{\mathrm{curr}}$ against $\mathcal{T}_{s,a,b_a(AC_{\mathrm{curr}})}$ of related actuators.
Thus, the same S-A mechanism supports both post-hoc diagnosis and robust handling of unseen actuator combinations.

\subsubsection{System-level decision}
SCDT supports system-level monitoring by aggregating sensor-level decisions. Under an OR policy, a time window is flagged anomalous if any monitored sensor window is judged anomalous. This matches operational practice where any abnormal process behavior warrants an alert, while explanations are provided at the sensor and context level to support rapid diagnosis.

\section{Experiments}

All experiments were performed on the Intel Core i9-10900F CPU @ 2.80~GHz and the NVIDIA RTX~3080 GPU with CUDA~12.2. 
The RAPIDS cuML~25.12 toolkit was used to accelerate UMAP and HDBSCAN on GPU; CPU-based clustering was prohibitively slow for the scale of our per-S-AC window groups, whereas the GPU versions provide more than an order of magnitude speedup in practice. We use GPT-5-mini for both the Primary and Secondary LLM components. We set the Primary LLM to low verbosity with a maximum of 450 output tokens per call, while the Secondary LLM uses the same model without an explicit output token cap. All features are computed on fixed-length windows with $W=30$ samples.

\subsection{Datasets}

\begin{table}[ht]
    \centering
    \caption{Dataset Characteristics}
    \label{dataset}
    \begin{tabular}{lccc}
    \toprule Dataset & SWaT & WADI & HAI \\
    \midrule \#Train/Test & 1/1 & 1/1 & 4/2\\
    \#Process & 6 & 3 & 4\\
    \#Variables & 51 & 123 & 86\\
    \midrule
    \#Train time & 495,000 & 784,542 & 896,400 \\
    \#Test time & 449,920 & 172,800 & 284,400 \\
    Anomalies & 12.140\% & 5.778\% & 4.003\% \\
    \bottomrule
    \end{tabular}
\end{table}

We evaluate SCDT on three widely used public ICS datasets: SWaT~\cite{SWAT}, WADI~\cite{WADI}, and HAI~\cite{HAI}. 
Table~\ref{dataset} summarizes their characteristics.

\paragraph*{SWaT and WADI}
Both datasets use real industrial equipment for water treatment and distribution. WADI is physically interconnected with SWaT and can receive treated water (e.g., RO permeate) from the SWaT testbed as an inflow, making SWaT effectively upstream for WADI. SWaT contains 7 days of normal data and 4 days of attack data with 51 variables, while WADI extends the distribution process with 14 days of normal operation, 2 days of attack scenarios, and 123 variables. These datasets are standard benchmarks for ICS anomaly detection.

\paragraph*{HAI}
The HAI dataset integrates hardware-in-the-loop simulation with boiler, turbine, water-treatment modules and synthetic power-generation dynamics. HAI provides four independent training segments and two disjoint test segments with 50 attack scenarios. Its multi-process interactions, and segmented temporal structure make it a challenging benchmark requiring cross-segment generalization.

\subsection{Baseline Methods}

As summarized in Related Work, we compare SCDT with nine widely used MTS anomaly detectors spanning
classical distance/subspace methods \cite{PCA, KNN, OCSVM}, reconstruction-based models \cite{AE, DAGMM, LSTMVAE, MADGAN, USAD},
and structure-aware forecasting approaches \cite{GDN,FuSAGNet}.
Table~\ref{analysis} reports precision, recall, and F1 under the standard evaluation protocol used in prior ICS studies.

\begin{table*}[ht]
  \centering
  \caption{Comparison with baselines on SWaT, WADI, and HAI in terms of precision(Pre) (\%), recall (Rec) (\%), and f1-score (F1) (\%). The best performances for each evaluation metric are highlighted in bold and the second best are underlined. Baseline Results are partly from the work of Han \& Woo ~\cite{FuSAGNet}, and Deng \& Hooi ~\cite{GDN}}
  \label{analysis}
  \begin{tabular}{lccccccccc}
    \toprule
    Method 
      & \multicolumn{3}{c}{SWaT} 
      & \multicolumn{3}{c}{WADI}
      & \multicolumn{3}{c}{HAI} \\
    \cmidrule(lr){2-4}
    \cmidrule(lr){5-7}
    \cmidrule(lr){8-10}
      & Pre & Rec & F1 
      & Pre & Rec & F1 
      & Pre & Rec & F1 \\
    \midrule
    PCA & 24.92 & 21.63 & 23.16 & 39.53 & 5.63 & 9.86 & 37.44 & 12.09 & 18.28 \\
    KNN & 7.83 & 7.83 & 7.83 & 7.76 & 7.75 & 7.75 & 12.53 & 11.60 & 12.05 \\
    OCSVM & 62.88 & 67.33 & 65.03 & 5.77 & \textbf{100.0} & 10.91 & 3.76 & 34.92 & 6.79 \\
    AE & 72.63 & 52.63 & 61.03 & 34.35 & 34.35 & 34.35 & 53.55 & \underline{73.38} & 61.92 \\
    DAGMM & 27.46 & 69.52 & 39.37 & 54.44 & 26.99 & 36.09 & 21.38 & 68.79 & 32.62 \\
    LSTM-VAE & 96.24 & 59.91 & 73.85 & 87.79 & 14.45 & 24.82 & 57.83 & 70.51 & 63.54 \\
    MAD-GAN & 98.97 & 63.74 & 77.54 & 41.44 & 33.92 & 37.30 & 47.70 & 32.83 & 38.89 \\
    USAD & \textbf{100.0} & 56.00 & 71.79 & 43.09 & 22.51 & 29.57 & 9.32 & 13.35 & 10.98 \\
    GDN & \underline{99.35} & 68.12 & 80.82 & \textbf{97.50} & 40.19 & 56.92 & \underline{88.37} & 60.32 & 71.70 \\
    FuSAGNet & 98.78 & \underline{72.60} & \underline{83.69} & 82.92 & \underline{47.87} & \textbf{60.70} & 86.79 & \textbf{74.79} & \textbf{80.34} \\
    \midrule
    \textbf{SCDT} & 96.21 & \textbf{83.09} & \textbf{89.17} & \underline{89.68} & 43.13 & \underline{58.25} & \textbf{90.07} & 61.79 & \underline{73.30} \\
    \bottomrule
  \end{tabular}
\end{table*}

\begin{figure*}[ht]
  \centering
  \includegraphics[width=\linewidth]{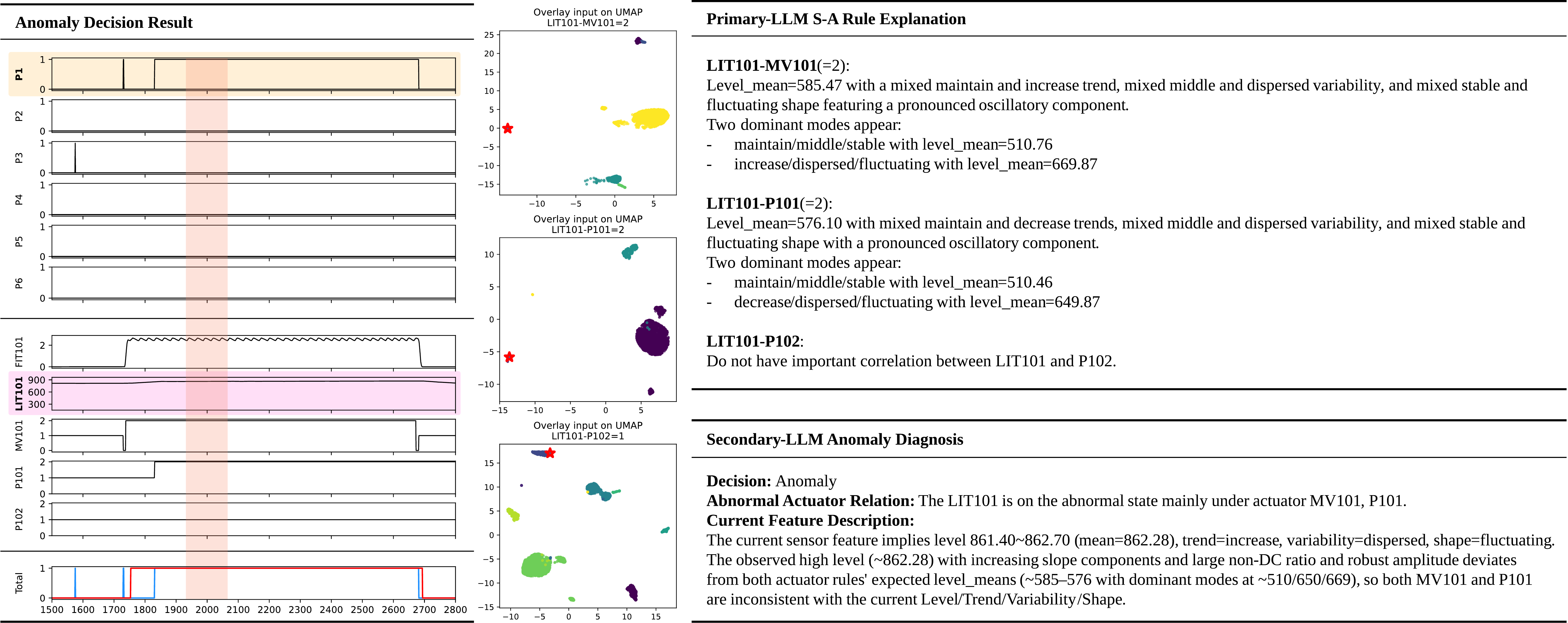}
  \caption{Example overview of anomaly diagnosis with LLM. The left panel shows anomaly detection results and device states. The middle plots project normal-window features of LIT101 conditioned on actuators using UMAP. The right tables report the Primary and Secondary LLM outputs.}
  \label{fig:llm_explanation}
\end{figure*}

\subsection{Evaluation Metrics}
We adopt precision, recall, and F1-score, which are the standard metrics in ICS anomaly detection. 
If an anomaly occurs at least once within the sampled detection window, we label that sample as anomalous.

\subsection{Performance Analysis}
\paragraph*{Performance Comparison with Baselines.} Table~\ref{analysis} reports precision, recall, and F1-score of SCDT against representative baselines on SWaT, WADI, and HAI. Although the primary contribution of SCDT lies in LLM-augmented, system-aware diagnosis rather than purely maximizing detection scores, SCDT remains highly competitive across all three benchmarks. In particular, SCDT achieves the best recall and F1-score on SWaT (Rec.=83.09\%, F1=89.17\%), indicating strong sensitivity to abnormal process behaviors under relatively stable actuator contexts. On WADI and HAI, SCDT attains the top-tier precision (WADI: 89.68\% as the second-best; HAI: 90.07\% as the best), while maintaining the second-best F1-score on both datasets (WADI: 58.25\%, HAI: 73.30\%). These results suggest that SCDT tends to detect anomalies more conservatively than some high-recall baselines, yielding fewer false alarms while still preserving robust overall detection quality.

\paragraph*{Precision–Recall Trade-off and Dataset-specific Behavior.}
We emphasize that performance should be interpreted in a balanced manner, particularly through the lens of F1-score, which jointly reflects false positives and false negatives. On WADI, for instance, some baselines achieve very high recall but at the cost of substantially reduced precision, which can increase the false-alarm burden in practice. In contrast, SCDT maintains top-tier precision on WADI and HAI while achieving the second-best F1-score on both datasets, indicating a more balanced operating point. The relatively lower recall of SCDT on WADI/HAI compared to SWaT can be attributed to actuator-context partitioning: certain actuator-key variables exhibit wide operating ranges, which creates a large number of distinct (and often sparse) S-AC contexts. This over-fragmentation reduces the number of normal windows per context, making multi-mode clustering and quantile-envelope rule estimation less statistically stable and increasing the likelihood of missed detections. Importantly, the same mechanism also explains the strong precision of SCDT, as alarms are triggered mainly for sensors with clear and consistent deviations supported by the available context rules.

\subsection{Anomaly Reasoning and LLM Explanations}

A representative true-positive anomaly diagnosis produced by our framework on the SWaT dataset is shown in Figure~\ref{fig:llm_explanation}. The left panel summarizes the anomaly detection results. The top six plots report the process-level detection scores, where SCDT identifies prominent anomaly peaks around 1600s and 1750s in processes P3 and P1, and subsequently flags an anomalous interval in P1 spanning approximately 1830-2700s. 
For clarity, the plot corresponding to Process~1 is highlighted in yellow, since the subsequent analysis focuses on a sensor-level explanation within P1. 
The next six plots show the raw trajectories of the sensors and actuators in P1, and the LIT101 trajectory is highlighted in pink as the primary diagnostic target in this case study.
We additionally mark the reference time point at 2000s in coral to indicate the specific window whose feature vector is projected and analyzed in the subsequent UMAP-based visualization. 
In the bottom plot, the red curve denotes the ground-truth anomaly label, indicating that the interval from roughly 1750s to 2700s is annotated as an attack. The blue regions correspond to the intervals detected as anomalous by SCDT across processes; the detected window aligns well with the labeled attack period, demonstrating that the method captures the majority of the abnormal segment.

The middle three plots visualize the distributions of normal-window features for LIT101 with respect to MV101, P101, and P102, projected onto a two-dimensional manifold using UMAP. Different colors indicate distinct normal rules inferred by the Primary LLM. The red star marks the feature vector of LIT101 at time 2000s after projection into the same UMAP space. As shown, the projected point lies far from the normal clusters for MV101 and P101, whereas it overlaps the normal region for P102. This contrast suggests that the abnormality at this time is most inconsistent with the operating regimes associated with MV101 and P101, while remaining compatible with the feature distribution observed under P102.

The two tables on the right report the outputs of the Primary and Secondary LLMs, respectively. The Primary LLM characterizes the relationship between LIT101 and the current actuator state, and summarizes each S-A rule in one to two sentences. Consistent with the UMAP visualizations, the rules for MV101 and P101 capture distinct normal regimes that explain the structure of the feature distributions. For P102, the relation is indicated as weak or non-informative; this occurs when the conditional distribution of LIT101 under the current P102 state is effectively indistinguishable from the marginal distribution across all actuator states, preventing a meaningful separation into actuator-specific regimes.

The Secondary LLM performs context-aware reasoning at inference time. Given the Primary LLM's rule set for sensor-actuator relations, it (i) summarizes the current window by inferring its Level, Trend, Variability, and Shape from the observed features, (ii) compares this description against the relevant normal rules conditioned on the current actuator status, (iii) identifies the actuator relations that exhibit the strongest inconsistency, and (iv) produces a final anomaly decision with a concise explanation. In the example, the current feature description on time 2000s indicates an abnormal water-tank level characterized by an increasing and unstable pattern, and the inferred abnormal relations highlight MV101 and P101 as the most plausible contributors.

This interpretation is consistent with the SWaT attack log: during the labeled attack interval (approximately 1750-2700s), MV-101 is compromised. Specifically, MV-101 is forced open when it should remain closed, leading to tank overflow and corresponding abnormal behavior in LIT101. Through the Secondary LLM’s reasoning, the diagnosis not only detects the anomalous interval but also provides actionable guidance by indicating that MV101 and P101 should be examined as primary suspects in the incident chain.

\section{Conclusion}
We proposed SCDT, a system-aware unsupervised framework for ICS anomaly diagnosis that integrates context-conditioned detection with LLM-grounded explanations. SCDT learns normal behavioral envelopes under actuator contexts via multi-stage S--AC clustering and supports fast online screening that reflects the context dependence of closed-loop control. Experiments on SWaT, WADI, and HAI showed that SCDT achieves competitive detection performance relative to representative multivariate time-series baselines while avoiding reliance on labeled attack data.

SCDT also improves diagnostic usability. By converting cluster-level feature statistics and sensor--actuator rules into concise semantic expectations, the Primary/Secondary LLM pipeline helps connect observed deviations to plausible actuator-related causes and operator-facing verification steps. The qualitative analysis indicates that the learned rule bank captures meaningful operational relationships and yields consistent, evidence-grounded explanations rather than opaque anomaly scores.

The proposed reasoning module is modular by design. Since it operates on structured evidence such as extracted features, matched contexts, deviation margins, and retrieved rule texts, it can be attached as a diagnostic layer on top of alternative detectors. This enables a practical pathway to pair stronger detection backbones with actuator-aware, context-grounded explanations, improving interpretability and operational utility in real industrial deployments.

\bibliographystyle{unsrtnat}
\bibliography{references}  






\end{document}